\documentclass[english,prl,twocolumn,superscriptaddress,preprintnumbers,amsmath,amssymb,showpacs]{revtex4}
\usepackage[T1]{fontenc}
\usepackage[latin1]{inputenc}
\usepackage{graphicx}
\newcommand{\beq}{\begin{equation}}
\newcommand{\eneq}{\end{equation}}
\newcommand{\beqy}{\begin{eqnarray}}
\newcommand{\eneqy}{\end{eqnarray}}
\makeatletter
\usepackage{amsmath}
\usepackage{epsfig}
\usepackage{dcolumn}
\usepackage{bm}
\usepackage{rotating}

\input{epsf}
\usepackage{color}
\usepackage{graphicx}
\usepackage{mathrsfs}
\usepackage{theorem}
\usepackage{amsmath,amssymb}

\newcommand{\ket}[1]{\left| #1 \right\rangle}

\usepackage{dcolumn}
\usepackage{bm}
\usepackage{times}

\usepackage{multirow}

\def\ket#1{\mathinner{|{#1}\rangle}}

\usepackage{babel}
\makeatother
\begin{document}

\title{Giant-Kerr nonlinearities in Circuit-QED}
\author {Stojan Rebi\'{c}}\author{Jason Twamley}
\affiliation{Centre for Quantum Computer Technology, Physics Department, Macquarie University, Sydney, NSW 2109, Australia}
\author{Gerard  J. Milburn}
\affiliation{Centre for Quantum Computer Technology, Department of Physics, University of Queensland, St Lucia, QLD 4072, Australia}

\begin{abstract}
The very small size of optical nonlinearities places strict restrictions on the types of novel physics one can explore. For an ensemble of multilevel systems one can synthesize a large effective optical nonlinearity using quantum coherence effects but such non-linearities are technically extremely challenging to demonstrate at the single atom level. This work describes how a single artificial multi-level Cooper Pair Box {\em molecule},  interacting with a superconducting microwave coplanar resonator, when suitably driven, can generate extremely large optical nonlinearities at microwave frequencies, with no associated absorption. We describe how the giant self-Kerr effect can be detected by measuring the second-order correlation function and quadrature squeezing spectrum.
\end{abstract}

\maketitle
%\indent 

{\em Introduction:-} Given a sufficiently large optical nonlinearity with low quantum noise, it should be possible to generate and observe strictly quantum effects in electromagnetic fields. Examples of such effects include quadrature squeezing~\cite{WallsMilburn}, generation of a superposition of macroscopically distinct quantum states~\cite{Yurke86}, optical switching with single photons~\cite{Dayan08} and measurements of nonlocal correlations of entangled photon states~\cite{Ou88}. So far, the successful demonstration of these effects has been limited to implementations with photons and atoms. The main obstacle for such an implementation -spontaneous emission- can be bypassed by exploiting quantum coherence effects in multilevel atoms. Such effects include coherent population trapping (CPT) ~\cite{Arimondo96}, electromagnetically induced transparency (EIT)~\cite{Fleischhauer05} and others. 

Recently, a novel system was shown to be capable of implementing basic quantum optical systems. Circuit quantum electrodynamics (cQED) is an on-the-chip counterpart of cavity QED systems~\cite{Blais04}, that employs a quantised microwave mode held in a Co-Planar Waveguide resonator (CPW) (substituting the standing-wave optical cavity) and a Cooper Pair Box (CPB) (instead of two-level atom trapped in the cavity). This system offers an unprecedented level of tunability and flexibility in the implementation of strong-coupling interaction limit.
\begin{figure}[t]
\begin{center}
\setlength{\unitlength}{1cm}
\begin{picture}(4,4.5)
\put(-1.25,0){\includegraphics[width=7.25cm,height=4.5cm]{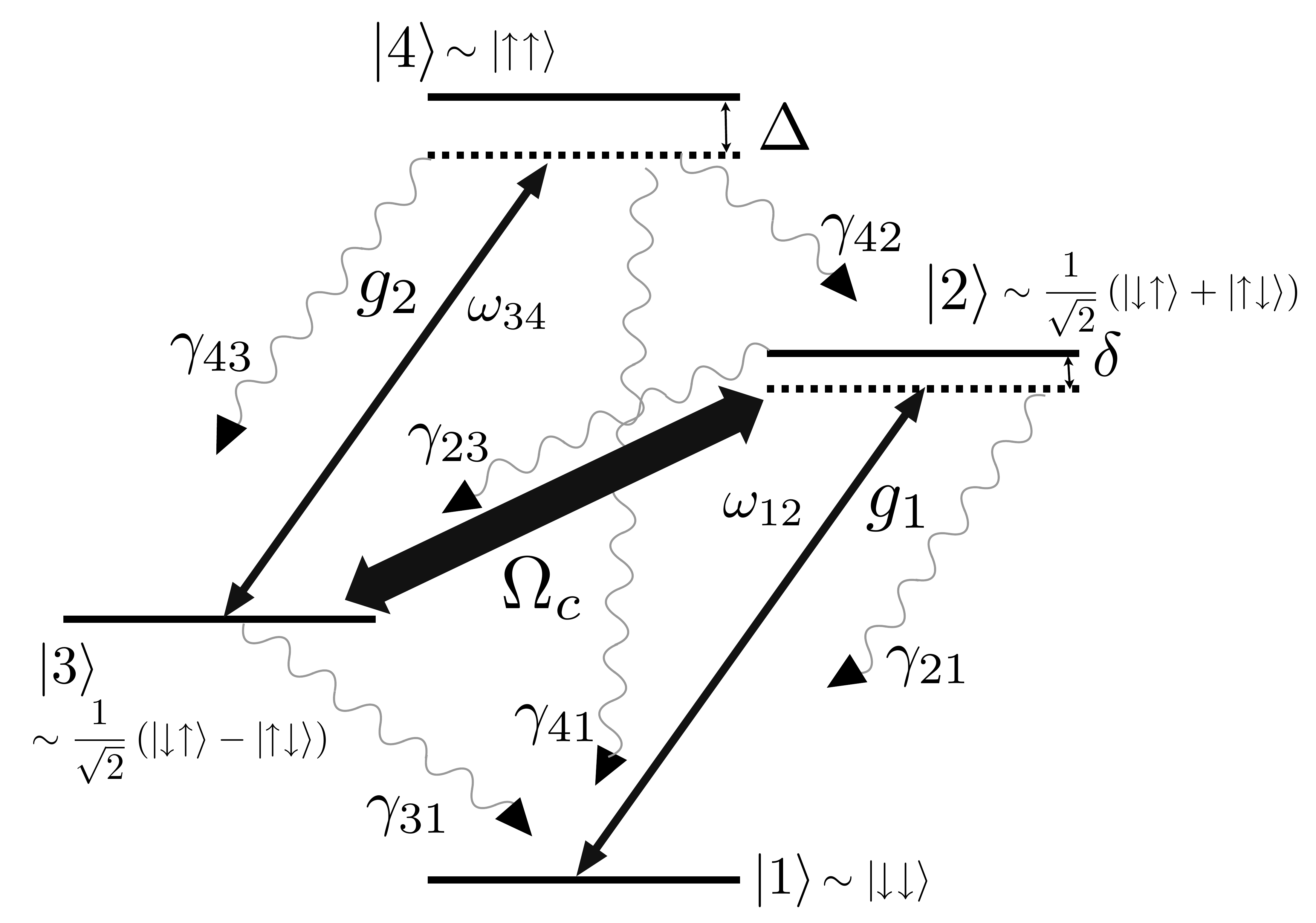}}
\end{picture}
\end{center}
\vspace{-0.5cm}
	\caption{N-system \cite{Schmidt96}, constructed from two coupled two-level CPB systems where transitions $|1\rangle \leftrightarrow |2\rangle$ and $|3\rangle\leftrightarrow|4\rangle$ are coupled with strengths $g_1,\;g_2$, to the microwave photonic mode $\hat{a}$ of frequency $\omega_a$, held in the CPW resonator. The transition $|2\rangle\leftrightarrow|3\rangle$ is driven by the semiclassical control field $\Omega_c$, while $\gamma_{ij}$, denotes the decay rate from $|i\rangle\rightarrow|j\rangle$. We define $\Delta = \omega_{34} - \omega_a$ and $\delta = \omega_{12} - \omega_a$.
	%% insert definition of deltas here
	}
\label{N-System}
\vspace{-0.5cm}
\end{figure}
The aim of this paper is to show that, by designing and  utilizing multi-level coherent processes in ``artificial superconducting atoms'',  Circuit-QED systems can display effects completely analogous to CPT and EIT. In this work we show that by designing a specific type of four level ``artificial superconducting atom'', and arranging for EIT in this system when coupled to the CPW, one obtains a giant Kerr nonlinearity (Fig.~\ref{N-System}). EIT is based on the use of dark resonance where quantum interference cancels the absorption of the probe signal. A strongly detuned fourth level provides an ac-Stark shift to the ground state $|3\rangle$, resulting in a self-Kerr nonlinearity free of spontaneous emission noise. Schmidt and Imamo\u{g}lu \cite{Schmidt96} predicted that this $N$-scheme can give rise to several orders of magnitude enhancement in Kerr nonlinearity as compared to conventional schemes. Their prediction has been verified in a recent experiment by Kang and Zhu \cite{Kang03}, although in a semiclassical regime. In our work we show that, due to the very strong coupling between the ``artificial superconducting atom'' and the CPW, the resulting giant Kerr nonlinearity is predicted to be three-to-four orders of magnitude larger than what has been so far experimentally demonstrated. Moreover we predict that the effect could be quite robust against dephasing. The presence of such large Kerr nonlinearities in a high-finesse cavity could result in complete photon blockade giving an effective two-level behavior for the cavity mode~\cite{Imamoglu97,Gheri99,Rebic99,Werner99,Rebic02}. A large effective self-Kerr nonlinearity allows for a photon blockade effect, i.e. a source of well resolved single photons. It could also be potentially adopted for use as a microwave single photon detector and to implement conditional quantum logic. 

Below we show that one way to tailor the required multi-level ``artificial superconducting atom'', giving a four-level $N$-scheme (4 atomic levels with transitions in the shape of the letter $N$), is via the straightforward capacitive coupling of two CPBs. When this system is coupled to a quantized CPW field, severe photon blockade can be observed, corresponding to a huge nonlinear photon-photon interaction. The effect of quadrature squeezing is also predicted. Together, the Giant-Kerr effect of the N-scheme, with the very strong coupling in cQED (due to the large dipole moment of the CPB and the small mode volume of the CPW), gives the posibility of immensely large optical nonlinearities at the single photon level. 

{\em CPB Molecule:-} Normally CPBs are operated at the charge degeneracy point and act as an effective two level system when the charging energy greatly exceeds the Josephson energy, i.e. $E_C\gg E_J$. To model a multilevel atomic system, one can capacitively couple two CPBs together to form a {\em CPB molecule}. For weak coupling, the CPB-molecule's states $|\!\!\!\uparrow\downarrow\rangle$, $|\!\!\!\downarrow\uparrow\rangle$ are nearly degenerate, while for large coupling the corresponding eigenstates  are non-perturbative superpositions of these bare states and are strongly split. This results in a formation of multi-level system~\cite{Blais07}. For  zero detunings $\Delta=\delta=0$, we arrange for resonance between the $34$, $12$ transitions and the cavity, $\omega_{34}=\omega_{12}=\omega_a$. We now show that by operating the capacitively coupled CPBs at the co-resonance point, a symmetric $N-$system is realised (see Fig. \ref{N-System}). With this design we also have the important flexibility to adjust the level structure: by tuning the flux threading the CPBs equally within the {\em CPB-molecule}, an asymmetry is introduced leading to nonzero detunings $\Delta,\;\delta$.

\begin{figure}[t]
\vspace{-0.5cm}
\begin{center}
\setlength{\unitlength}{1cm}
\begin{picture}(4,3.5)
\put(-2.,0){\includegraphics[width=8cm,height=3cm]{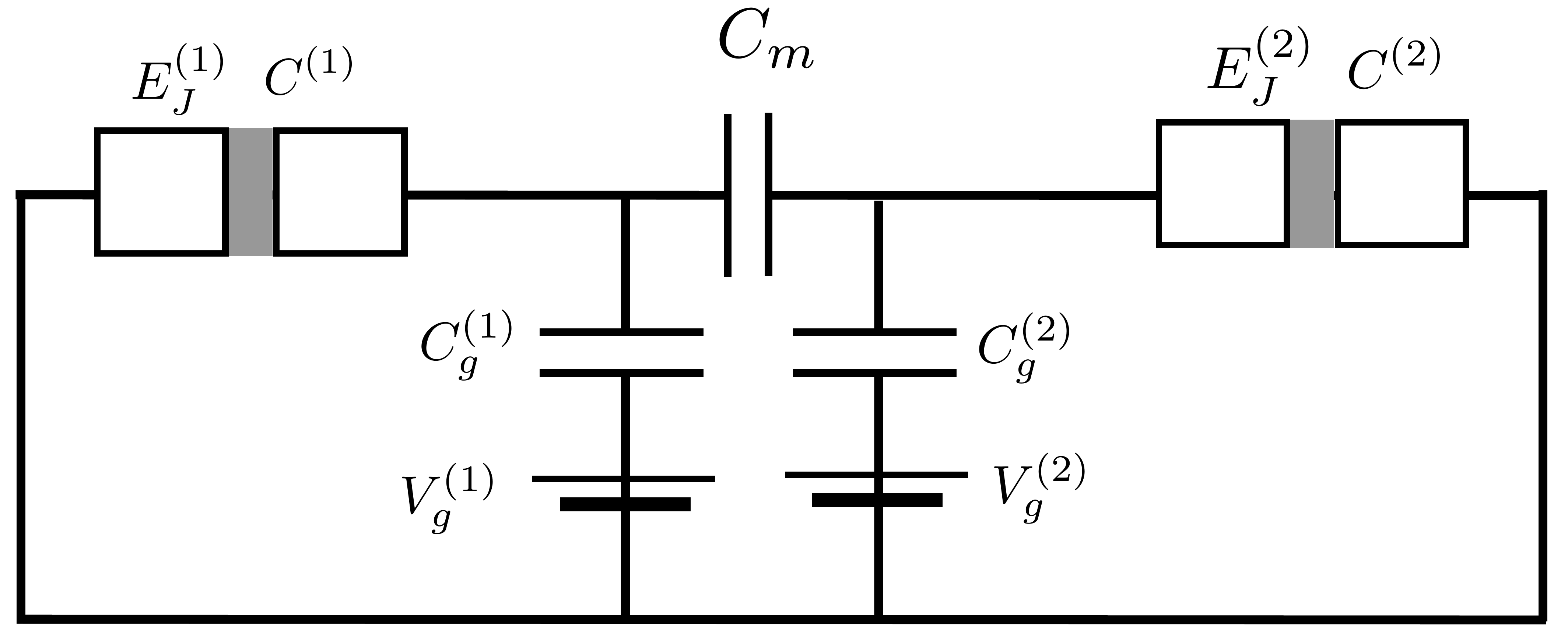}}
\put(-.75,-4){\includegraphics[scale=0.16]{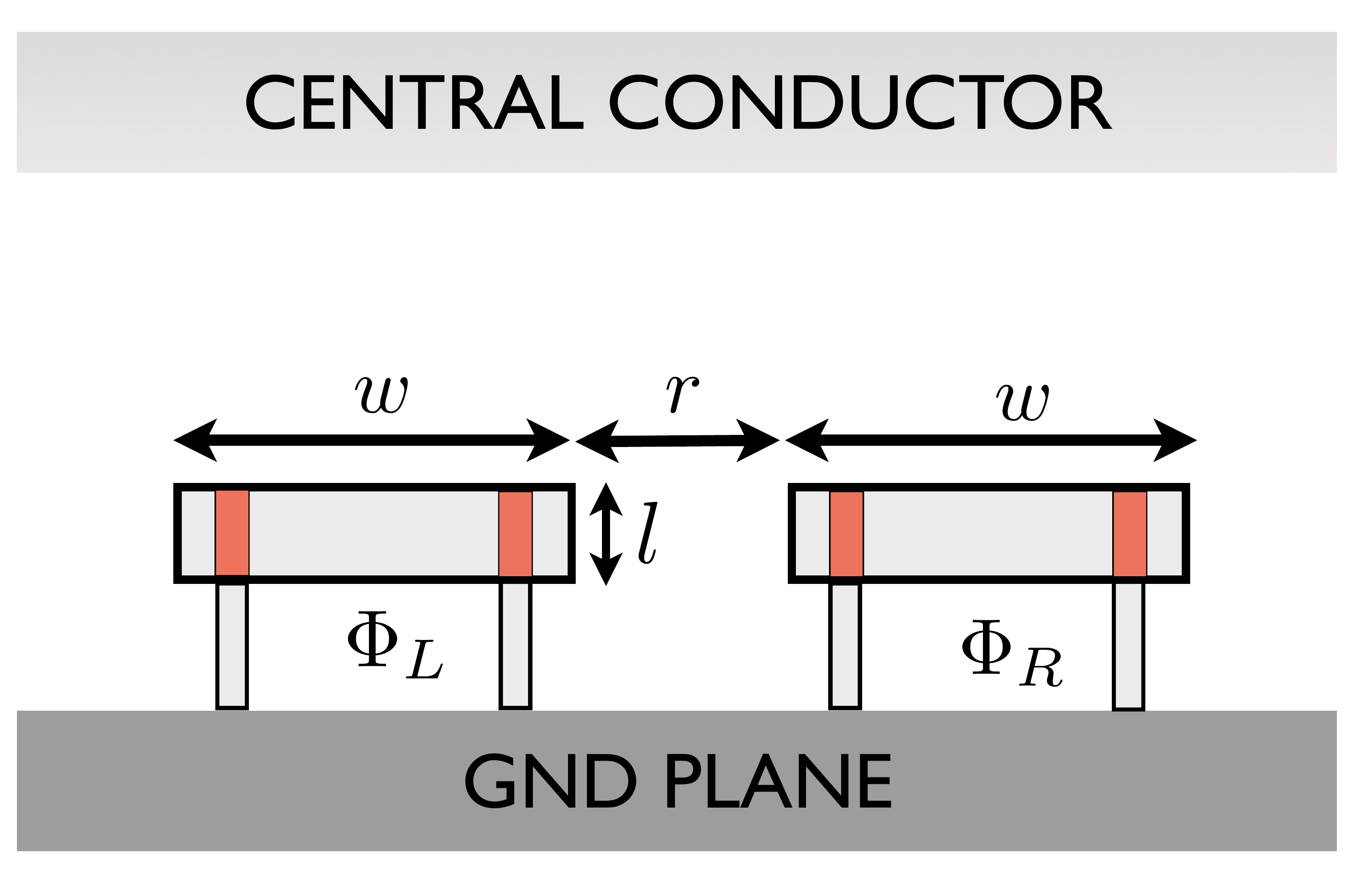}}
\put(0.35,2.5){\large(A)}
\put(-1.55,-0.65){\large(B)}
\end{picture}
\end{center}
\vspace{3.5cm}
	\caption{(A) Circuit of two capacitively coupled CPBs with individual gate bias; (B) Schematic of possible physical arrangement of an identical ($\Phi_L=\Phi_R$), pair of CPBs to yield a {\em CPB molecule}.}
\label{Circuit}
\vspace{-0.25cm}
\end{figure}

Considering two capacitively coupled CPBs  (Fig.~\ref{Circuit}(A)), the associated Hamiltonian is a combination of ``Kinetic'' (KE) and ``Potential'' (PE) terms associated with the phase difference of the wavefunctions across each Josephson junction $\phi_{1(2)}$. From the properties of  Josephson junctions one has $\dot{\phi}_j=2eV_j/\hbar$,  where $V_j$ is the voltage drop across the $j^{\textrm{th}}$ Josephson junction, and the current through the junction is $I_j=I_c\sin\phi_j$. The total energy required to charge all the capacitors in the circuit gives $KE=1/2\sum_{j=1}^2\,(C^{(j)}V^{(j)\,2}+C_{g}^{(j)}V_{g}^{(j)\,2})+1/2C^{(m)}V^{(m)\,2}$, while the total $PE=\sum_{k=1}^2{\cal E}_{J}^{(k)}(1-\cos\phi^{(k)})$. ${\cal E}_{J}^{(k)}$ is the Josephson energy of the $k^{\textrm{th}}$ junction. Starting from Kirchoff's laws, we obtain the classical Hamiltonian
%Using Kirchoff's laws one can solve for $V_{g}^{(j)}(V^{(1)},V^{(2)})$, and $V^{(m)}(V^{(1)},V^{(2)})$, and thus form the equivalent classical Lagrangian ${\cal L}(\dot{\phi}^{(j)},\phi^{(j)})$, derive the canonical momentum $p^{(j)}=\hbar n^{(j)}\equiv {\cal L}_{\dot{\phi^{(j)}}}$, perform a Legendre transformation, to find the equivalent Hamiltonian, ${\cal H}=\sum_j\,p^{(j)}\dot{\phi}^{(j)}-{\cal L}=\hbar\sum_j n^{(j)}\dot{\phi}^{(j)}-{\cal L}$, 
\begin{eqnarray}
H&=&\sum_{j=1}^2\,\left(E_c^{(j)}(n^{(j)}-N_{g}^{(j)})^2+{\cal E}_J^{(j)}(1-\cos\phi^{(j)})\right)\nonumber\\
&\ &+E_m(n^{(1)}-N_{g}^{(1)})(n^{(2)}-N_{g}^{(2)})\;\;,\label{Hclassical}
\end{eqnarray}
where the following quantities have been defined
$\{E_C^{(j)},E_m\}=0.5(2e)^2\{(C_{eff}^{(j)})^{-1},(C_{eff}^{(m)})^{-1}\}$,
$\{C_{eff}^{(j)},C_{eff}^{(m)}\}=\Xi\{(C_{\sum_{\{1,2\}\backslash\{j\}}})^{-1},(C^{(m)})^{-1}\}$, $\Xi= C_{\sum_1}C_{\sum_2}-C^{(m)2}$, %and
%\begin{equation}
$C_{\sum_j}=C^{(m)}+C_{g}^{(j)}+C^{(j)}$,
%\end{equation}
%%\begin{subequations}
%%\label{Eq:HamCoeffs}
%%\begin{eqnarray}
%E_C^{(j)}&=& \frac{(2e)^2}{2C_{eff}^{(j)}}\;\;,\;\;\; E_m=\frac{(2e)^2}{2C_{eff}^{(m)}}\;\;,\label{eq1} \\
%%\end{equation}
%%\begin{equation}
%C_{eff}^{(j)} &=& \frac{\Xi}{C_{\sum_{\{1,2\}\backslash\{j\}}}}\;\;,\;\;\;  C_{eff}^{(m)}=\frac{\Xi}{C^{(m)}}\;\;,\label{eq2} \\
%C_{\sum_j}&=&C^{(m)}+C_{g}^{(j)}+C^{(j)}, \; \Xi= C_{\sum_1}C_{\sum_2}-C^{(m)2}\;\;,\label{eq3}
%\end{eqnarray}
%\end{subequations}
and where the number of excess Cooper Pairs on the gates is $N_{g}^{(j)}=-C_g^{(j)}V^{(j)}/(2e)$. We consider the quantised version of (\ref{Hclassical}) %i.e. $H\rightarrow\hat{H},\; n^{(j)}\rightarrow\hat{n}^{(j)}$, 
and focus on the low energy dynamics by only including $|n^{(1)},n^{(2)}\rangle,\;\;n^{(j)}=1,2$ states in an expansion of the Hamiltonian operator. Setting $\delta^{(j)}\equiv N_g^{(j)}-1/2$, and denoting $\hat{Z}=|0\rangle\langle 0|-|1\rangle\langle 1|$, $\hat{X}=|0\rangle\langle 1|+|1\rangle\langle 0|$ gives
\begin{equation}
\hat{H}/\hbar=\sum_{j=1}^2\,\left(\omega_z^{(j)}\delta^{(j)}\hat{Z}^{(j)}-\omega_x^{(j)}\hat{X}^{(j)}\right)+J\hat{Z}^{(1)}\hat{Z}^{(2)}\;\;,\label{quantHam}
\end{equation}
where $\hbar\omega_z^{(j)}=E_C^{(j)}+E_m/2$, $\hbar\omega_x^{(j)}={\cal E}_J^{(j)}/2$, and $\hbar J=E_m/4$. If the individual CPBs are  identical  then the individual Josephson energies in (\ref{quantHam}) can be modulated by local flux tuning via $\hbar\omega_x^{(j)}=E_J^{(j)}=2{\cal E}_J\cos(\pi \Phi^{(j)}/\Phi_0)$. Working at the co-
% corrected superscript
resonance point $\delta^{(j)}=0$, and setting $\omega_x^{(1)}=\omega_x^{(2)}=\omega_x$ yields
\begin{equation}
\hat{H}/\hbar=-\omega_x\left(\hat{X}^{(1)}+\hat{X}^{(2)}\right)+J\hat{Z}^{(1)}\hat{Z}^{(2)}\;\;.\label{quantHam1}
\end{equation}
The eigenenergies of (\ref{quantHam1}), labeled as in Fig.~\ref{N-System}, are $(E_4,E_2,E_3,E_1)=J\times(\epsilon,1,-1,-\epsilon)$, where $\epsilon^2 = 1 + 4\omega_x^2/J^2$. The transition frequencies are then $\omega_{21}=\omega_{42}=J(\epsilon+1)$ and $\omega_{23}=2J$. To arrange for nonzero detunings $\Delta\neq\delta$ (i.e., $R\equiv\omega_{42}/\omega_{21}\ne 1$), one must move slightly off the co-resonance point by introducing equal strength Zeeman terms in (\ref{quantHam}),
\begin{equation}
\hat{H}/\hbar=\sum_{j=1}^2\,\bar{\omega}_z^{(j)}\hat{Z}^{(j)}-\omega_x\left(\hat{X}^{(1)}+\hat{X}^{(2)}\right)+J\hat{Z}^{(1)}\hat{Z}^{(2)}\;\;,\label{quantHam2}
\end{equation}
with $\bar{\omega}_z^{(j)}=\omega_z^{(j)}\delta^{(j)}$. $R$ can be set to any value $R\ge1$ so adjusting $\bar{\omega}_z$ and $\omega_x$ fixes the detunings $\Delta$, $\delta$ to any desired value without individually addressing each CPB.

To estimate the size of the coupling $J$, consider a closely spaced pair of CPBs as shown in Fig.~\ref{Circuit}(B) ~\cite{Gywat06}. The capacitance between the two CPBs can be expressed as $C^{(m)}=2\pi r\epsilon\epsilon_0w/\xi$, where 
%\begin{equation}
$\xi=K(\frac{2r}{l})+K(\frac{2w}{l})-K(\frac{2(w+r)}{l})$, 
%\end{equation}
with $K(x)=x\sinh^{-1}(1/x)+\sinh^{-1}(x)$ and $\epsilon=9$ the relative permittivity of the substrate material~\cite{Mariantoni08}. We can estimate that if  $(l,w,r)=(50,10,0.5)\mu{\rm m}$ we obtain a large splitting: $(\omega_z^{(j)},\;J)=(14.7, 2.3)$GHz. 
\begin{table*}
\label{eta-table}
\begin{tabular}{l|c|c|c|r}\hline\hline
\multicolumn{1}{c|}{Work} & \multicolumn{1}{c|}{$g/2\pi$(MHz)} & \multicolumn{1}{c|}{$\kappa/2\pi$(MHz)} & \multicolumn{1}{c|}{$\gamma/2\pi$(MHz)} & \multicolumn{1}{c}{$\eta/\kappa$} \\ \hline\hline
D.\ Englund {\it et al.} Nature 450, 857 (2007) & 8000 & 16000 & 100 & 2 \\ 
P.\ Maunz, {\it et al.}, Nature 428, 50 (2004) & 16 & 1.4 & 3 & 3 \\ 
K.\ M.\ Birnbaum {\it et al.}, Nature 436, 87 (2005) & 33 & 4.1 & 2.5 & 5.4 \\ 
C.J.\ Hood {\it et al.}, PRL\ 80, 4157 (1998). & 120 & 40 & 2.6 & 6.9 \\ 
A.\ Imamo\={g}lu {\it et al.}, PRL\ 79, 1467 (1997) &  &  &  & 20 \\ 
CPB Molecule & 300 & 1 & 0.1 & 45,000 \\ 
\end{tabular}
\caption{Table of $\eta$ from Eqn (\ref{eta}), measuring the size of the effective self-Kerr nonlinearity for various  quantum systems coupled to cavities. We take the quoted values for $(g,\kappa,\gamma)$, and consider additional driving (see Fig. \ref{N-System}) such that $(g/\Omega_c)^2=0.1$, [condition of validity for Eqn (\ref{eta})], take $(\Delta,\delta)=(\gamma,0)$ and $\gamma_i=\gamma$. The effective self-Kerr nonlinearity for the CPB-molecule/CPW system has the potential to be extremely large. The actual model studied in this Letter contains extra decays in addition to those traditionally studied in $N-$systems.}
\vspace{-0.55cm}
\end{table*}
Due to the nonlinear relationship between $C^{(m)}$ and $\hbar J=E_m/4$, the frequency $\omega_{32}\sim 2J$ for the co-resonance case can be set to be MHz-GHz depending on the CPB geometry. The value ${\cal E}^{(1,2)}_{max}=2\omega_x^{(1,2)}\sim 8$ GHz \cite{Blais04} can be reduced via adjusting the flux threading both CPB (assuming identical CPBs $\Phi_L=\Phi_R$). As a prototype, we choose $\omega_{43}=\omega_{21}=\omega_a=5$ GHz, giving vanishing detunings $\Delta=\delta=0$, if one arranges to work at the co-resonance point $\delta^{(1,2)}=0$. Then by driving the $2\leftrightarrow 3$ transition with RF at frequency $\omega_{23}$, one gets $J=0.2$GHz, and $\omega_x^{(1,2)}={E}^{(1,2)}_J=2.6$ GHz. To arrange for non-vanishing detunings for example, we can choose to set $\omega_x/2\pi=4$GHz, $J/2\pi=0.2$GHz, $\omega_z^{(1)}=\omega_z^{(2)}=2\pi\times 16$GHz, and with $\kappa/2\pi=100$kHz, one can obtain significant differences between the transition frequencies $\bar{R}\equiv \omega_{34}-\omega_{12}$, to yield $\bar{R}/\kappa\sim 1$. One can achieve these moderate detunings even when the gate bias is offset from the precise co-resonance point by a tiny amount $\delta^{(1)}=\delta^{(2)}=2.8\times 10^{-3}$. In the special case of identical CPBs operating exactly at the co-resonance point one has selection rules which could alter the transition strengths within our system but as we work off co-resonance we expect all transitions and decay paths to be allowed. This is a significant departure from the atomic $N$-system studied in \cite{Schmidt96}, and we will include all these decay routes in our full model  below. % be an issue for our scheme as we work slightly off co-resonance. However our scheme thus has new decay paths which are typically not allowed in the are now allowed and which must be accounted for.
%. Below we will require $|\Delta|/\kappa>0$, and this is achieved either by utilising identical CPBs off co-resonance, or using asymmetric CPBs, but in these two cases there will be no well defined transition rules.  In essence we deduce  that capacitively coupled pair of CPBs can be operated to simulate a four-level system akin to an atomic system but with the two important differences: $(a)$ with non-zero detunings there are no selection rules and thus sufficiently non-resonant transitions must be engineered to ``turn off'' all undesired excitations, and $(b)$ level $\ket{3}$ is not stable and decays at the same rates as levels $\ket{2}$ and $\ket{4}$.

{\em Giant Kerr Nonlinearity:-} It was shown~\cite{Imamoglu97,Gheri99,Rebic99,Werner99} that an $N$-system similar to one described in Fig.~\ref{N-System} yields an effective Kerr nonlinearity. Due to the absence of selection rules, we consider all possible decay paths with rates $\gamma_{ij}$. Drummond and Walls~\cite{Drummond80} analysed the pure $\chi^{(3)}$ (third-order optical nonlinear susceptibility \cite{WallsMilburn}, Sect: 5.4), Hamiltonian $\hat{H}_{eff}=\hbar\eta\,\hat{a}^{\dagger\, 2}\hat{a}^2$ in the presence of a dissipative cavity and using this  we will estimate the size of the effective self-Kerr nonlinearity $\eta$ we observe in our effective $N$-system. 
% By comparison with their work, the strength of the effective self-Kerr nonlinearity $\eta$ will be estimated for multilevel system of Fig.~\ref{N-System}.
%The combined cavity-artifical atom system goes far of the resonance upon the absorption of a single photon, effectively blocking the absorption of a second photon. 
For large $\eta$, the system emits photons in an antibunched manner, with large waiting times between single photon emissions. To probe these effects, the second order-correlation function 
%\begin{equation}
$g^{(2)}(\tau)=\langle a^\dagger(t)a^\dagger(t+\tau)a(t+\tau)a(t)\rangle/\langle a^\dagger (t)a(t)\rangle |^2$,
%\label{g2}
%\end{equation}
will be calculated, in particular $g^{(2)}(0)$, for the weakly pumped cavity. This quantity is accepted as a good measure for photon blockade~\cite{Imamoglu97,Rebic99,Rebic02,Birnbaum05}, and allows for a direct comparison with the analytical expression obtained in~\cite{Drummond80}.  

Before examining the combined resonator-CPB molecule in detail we can find a rough estimate for the achievable nonlinearity $\eta$ using parameters from a recent optical turnstile experiments (see Table~\ref{eta-table}), taking \cite{Rebic99}
\begin{equation}
\eta=\left(\frac{g_1}{\Omega_c}\right)^2\left(\frac{g_2^2\Delta}{\gamma_{43}^2+\Delta^2}-\frac{g_1^2\delta}{(\gamma_{21}+\gamma_{23})^2+\delta^2}\right)\;\;,\label{eta}
\end{equation}
which holds in the limit of $(g_1/\Omega_c)^2\ll 1$~\cite{Gheri99}. %Below we will make this more precise using numerical modeling. 

For the resonator-CPB molecule system we set $\gamma_{21}=\gamma_{23}=\gamma_{43}=\gamma=1/T_1$, where $T_1$ is the lifetime of the single CPB excited state, and we take $T_1=10\mu$s, $\kappa=1$MHz, and choose $(\Delta, \delta, \Omega_c)=(10, 0, 1)\gamma$. We choose the cross-relaxation rates $\gamma_{42;31;41}$ to vanish (only for the purpose of this estimate) and $g_1=g_2=g=300\kappa$, to obtain the enormously large Giant self-Kerr strength $\eta\approx 10\times 10^9\kappa$. Although (\ref{eta}) only holds when $g_j/\Omega_C\ll 1$, this computation gives one the hint that the system (with all decay channels operational) may yield very large nonlinearities and in the next section we model the full system to numerically verify this.

{\em Numerical Investigations:-} Estimating the size of the effective nonlinearity $\eta$ when $g_j>\Omega_c$ is not straightforward as the adiabatic approximation is not permitted~\cite{Imamoglu97}. For that reason, we solve the master equation for the density operator $\rho$, given by $\dot{\rho}=\dot{\rho}_{sys} + \mathcal{L}\rho$, where
\begin{subequations}
\label{eq:mastereq}
\begin{eqnarray}
\dot{\rho}_{sys} &=& -i\Delta_{21} \, \left[ \sigma_{22},\rho \right] -i\Delta_{31} \, \left[ \sigma_{33},\rho \right] -i\Delta_{41} \, \left[ \sigma_{44},\rho \right] \nonumber \\
&\ & - i g_1 \, \left[ a^{\dagger}\sigma_{12}+\sigma_{21}a, \rho \right] - i g_2 \, \left[ a^{\dagger} \sigma_{34}+\sigma_{43}a, \rho \right]  \nonumber \\
&\ & - i \left[ \Omega_c^*\sigma_{32}+\sigma_{23}\Omega_c, \rho \right]  - i E_p \, \left[a+a^{\dagger},\rho \right] \, , \label{eq:msys}
\end{eqnarray}
and
$\mathcal{L}\rho =\kappa \mathcal{D}[a]\rho+ \sum_{(ij)}\gamma_{ij} \mathcal{D}[\sigma_{ji}]\rho$, where $\mathcal{D}[A]B\equiv 2ABA^\dagger-\{A^\dagger A,B\}$.
%\begin{eqnarray}
%\mathcal{L}\rho  &=& \kappa \, \left( 2a\rho a^{\dagger} - a^{\dagger}a\rho - \rho a^{\dagger}a \right) \nonumber \\ 
%&\ & + \sum_{(ij)}\gamma_{ij} \, \left( 2\sigma_{ji}\rho \sigma_{ij} - \sigma_{ij}\sigma_{ji}\rho - \rho \sigma_{ij}\sigma_{ji} \right)  \, .\label{eq:mdamping}
%\end{eqnarray}
\end{subequations}
\begin{figure}[h]
\begin{center}
\setlength{\unitlength}{1cm}
\begin{picture}(7,6.5)
\put(-.4,-.2){\includegraphics[width=7.6cm,height=7cm]{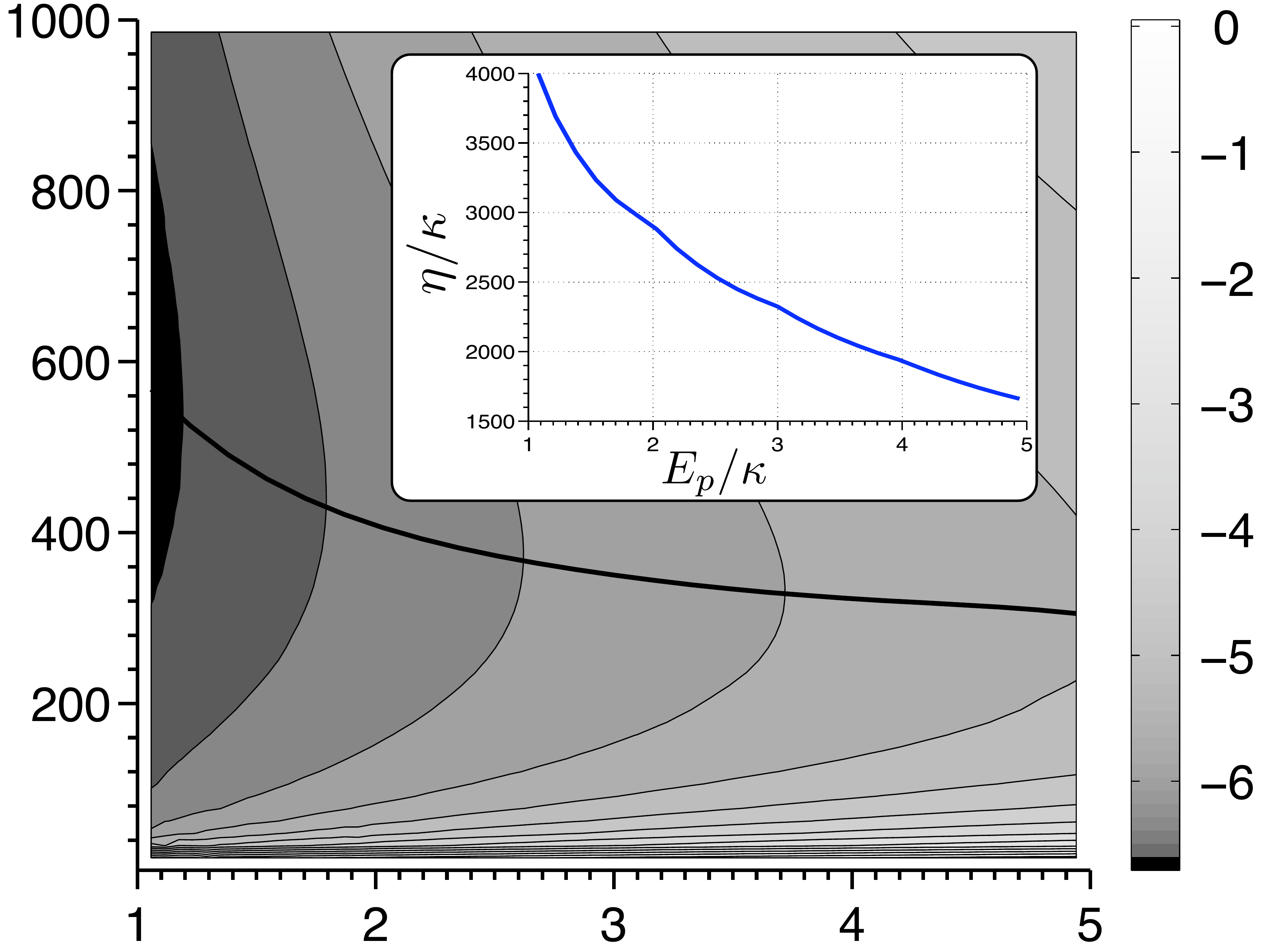}}
\put(3,-.5){\large$E_p/\kappa$}
\put(-.7,2.8){\begin{rotate}{90}\Large$\Omega_c/\kappa$\end{rotate}}
\end{picture}
\end{center}
	\caption{Graph of $\log_{10}[g^{(2)}(0)]$, for $g=300, \;\gamma_{23}=\gamma_{41}=.01$, and $\gamma_{21}=\gamma_{43}=\gamma_{31}=\gamma_{43}=.1$, in units of $\kappa$, and $(\Delta,\delta)=(.5,.5)$. For low $E_p$ and moderately large pumping $\Omega_c$, the autocorrelation drops to extremely low values $\sim10^{-6}$. The solid line indicates minimum $g^{(2)}(0)$. Inset shows the effective nonlinear coefficient $\eta$ vs. $E_p$ deduced from~\cite{Drummond80}, for the minimum $g^{(2)}(0)$ (see text for details). The average photon number is $\bar{n} \sim 10^{-2}$ for maximal value of the nonlinearity.}
\label{Autocorrelation}
\vspace{-0.45cm}
\end{figure}
Here, $\sigma_{ij}$ describes atomic transition operators $\hat{X}^{(ij)} = \sigma_{ij} + \sigma_{ji}$ for $i\neq j$, while $\sigma_{jj}$ models dephasing. $E_p \sim \sqrt{P\kappa/\omega_a}$ is the amplitude of an electric field driving the resonator mode, where $P$ is the power of the field incident on the resonator. Summation over $(ij)$ includes all decay channels shown in Fig.~\ref{N-System}. We solve $\dot{\rho}=0$ numerically to obtain $g^{(2)}(0)$. Fig.~\ref{Autocorrelation} shows the dependence on pump strength and (classical) coupling field strength. The correlation function increases with pump strength~\cite{Rebic02}. The dependence on the coupling field shows a decrease to a minimum (solid black line) followed by an increase. This is a consequence of a nonvanishing decay $\gamma_{31}$. Increasing $\Omega_c$ pumps the population in $\ket{1}$ and diminishes the effect of decay $\gamma_{31}$, hence the decrease in $g^{(2)}(0)$. Subsequent increase with $\Omega_c$ is due to a reduction in the nonlinearity~(\ref{eta}). Using the analytical result for $g^{(2)}(0)$ as a function of $E_p/\kappa$ in~\cite{Drummond80}, the effective self-Kerr nonlinearity $\eta/\kappa$ can be deduced (shown in the inset of Fig. 3). For weak driving, effective $\eta/\kappa \sim 10^3 - 10^4$ can be obtained. As highlighted in the introduction, using this "artificial $N$-system",  optical nonlinearities of unprecedented strengths are achievable using circuit QED. This result is robust against dephasing. When we include identical rates of dephasing on all levels (taking $\gamma_{kk} = \gamma_{ph} \sim 2.5\gamma_{ij}$, as seen in experiments when working off the critical point), we find that $\eta$ is an order of magnitude smaller than what is shown in Fig.~\ref{Autocorrelation}. Decreasing $E_p/\kappa$ further allows one to reach values shown in the Figure (and higher), and the use of transmons~\cite{Koch07} could reduce the dephasing rate when working off the critical point.

\begin{figure}[t]
\begin{center}
\setlength{\unitlength}{1cm}
\begin{picture}(7,6.4)
\put(-.4,-.3){\includegraphics[width=7cm,height=6.6cm]{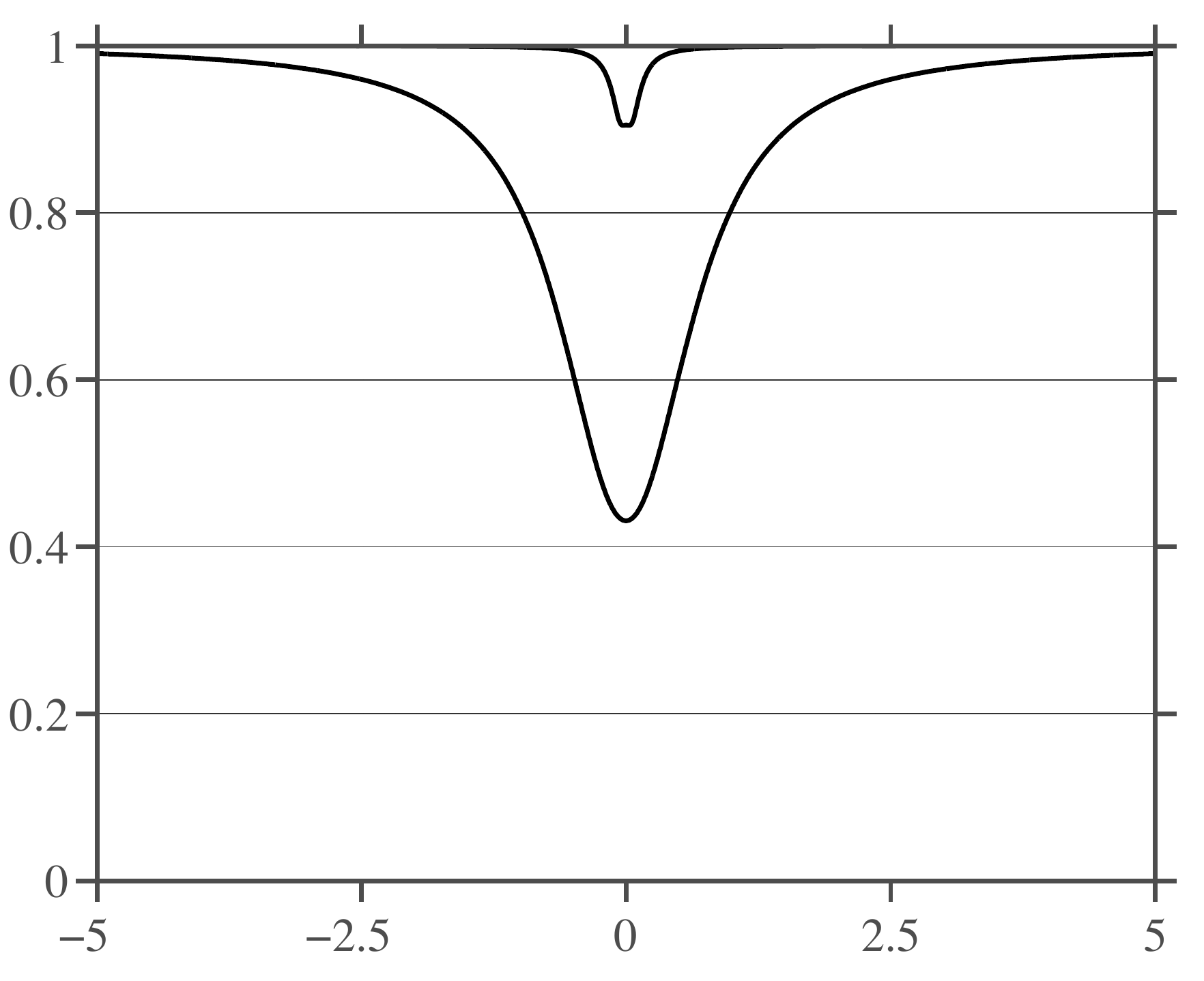}}
\put(3.3,-.6){\large$\omega/\kappa$}
\put(-.7,2.7){\begin{rotate}{90}\Large$S(\omega/\kappa)$\end{rotate}}
\put(3.9,2.5){(B)}
\put(3.8,5.){(A)}
\end{picture}
\end{center}
	\caption{Spectrum of squeezing $S(\omega/\kappa)$, for $(\Delta,\delta,g,E_p)=(5.13,-4.89,300,0.14)$ and (A): $\Omega_c=50$, or (B): $\Omega_c=1200$, and where $\gamma_{23}=\gamma_{41}=.01$ and $\gamma_{21}=\gamma_{43}=\gamma_{31}=\gamma_{43}=.1$, in units of $\kappa$. }
\label{Squeezing}
\vspace{-0.35cm}
\end{figure}
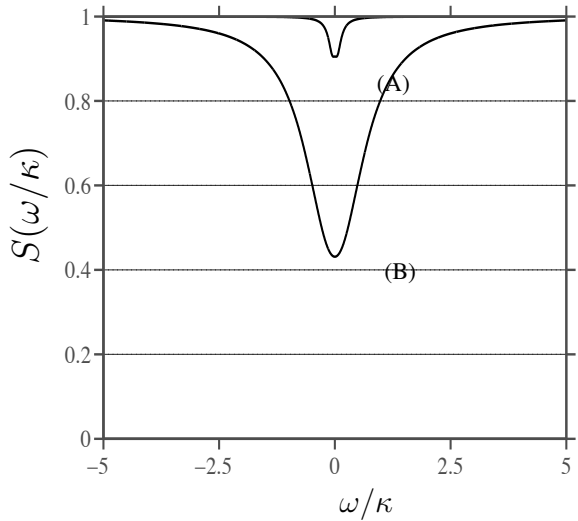

Self-Kerr systems can produce some degree quadrature squeezing~\cite{Collett85}. In Fig.~\ref{Squeezing}, the spectrum of maximum squeezing (belonging to the amplitude quadrature) is shown. The regime where maximum squeezing is obtained [Fig. 4(B)] is closer to the four-wave mixing regime. Given the presence of six coherence-destroying spontaneous decay channels, the result of Fig. 4 is remarkable, since pure dispersive Kerr medium can only achieve a squeezing of $\frac{2}{3}$~\cite{Collett85}.

While the measurement of squeezing is easy to implement using homodyne techniques, the measurement of second-order correlations involves efficient photon counters~\cite{Birnbaum05}, which are not currently available in the MW. We note, however, a recently technique~\cite{Grosse07} to obtainin $g^{(2)}(0)$ by homodyne methods. 

In conclusion, we have shown that an "artificial" multilevel system in circuit-QED, of two capacitively coupled CPBs coupled to the quantized field in a CPW, leads to the generation of giant self-Kerr nonlinearities as demonstrated by strong antibunching and quadrature squeezing. The size of the effective nonlinearity is orders of magnitude larger than anything previously known, owing to the combined effects of quantum coherence and strong coupling.

We acknowledge helpful discussions with T Duty and thank European Commission FP6 IST FET QIPC project QAP Contract No. 015848, DEST ISL Grant CG090188 and ARC (DP0986932) (SR \& JT).

\end{document}